\documentclass[12pt]{iopart}

\usepackage{iopams, bm, graphicx}  
\begin{document}

\newcommand{\be}{\begin{equation}}
\newcommand{\ee}{\end{equation}}
\newcommand{\bea}{\begin{eqnarray}}
\newcommand{\eea}{\end{eqnarray}}
\newcommand{\ad}{a^{\dag}}
\newcommand{\la}{\langle}
\newcommand{\ra}{\rangle}
\newcommand{\om}{\omega}
\newcommand{\bfr}{{\bf r}}

\title[Entanglement, complementarity, and vacuum fields]{Entanglement, complementarity, and vacuum fields in spontaneous parametric down-conversion}
\author{R Menzel$ˆ1$, A Heuer$ˆ1$,
P Milonni$ˆ2$,$ˆ3$}
\address{$ˆ1$ University of Potsdam, Institute of Physics and Astronomy, Karl-Liebknecht Stra\ss e 24-25,
D-14476 Potsdam, Germany}
\address{$ˆ2$ Theoretical Division, Los Alamos National Laboratory, Los Alamos, New Mexico 87545 USA}
\address{$ˆ3$ Department of Physics and Astronomy, University of Rochester, Rochester, NY 14627, USA}
\ead{menzel@uni-potsdam.de}

\begin{abstract}
Using two crystals for spontaneous parametric down-conversion in a parallel setup, we observe two-photon interference with high visibility. The high visibility is consistent with complementarity and the absence of which-path information. The observations are explained as effects of entanglement or equivalently in terms of interfering probability amplitudes, and also by the calculation of a second-order field correlation function in the Heisenberg picture. The latter approach brings out explicitly the role of the vacuum fields in the down-conversion at the crystals and in the photon coincidence counting. For comparison we show that the Hong-Ou-Mandel dip can be explained by the same approach in which the role of the vacuum signal and idler {\sl fields}, as opposed to entanglement involving vacuum {\sl states}, is emphasized.
\end{abstract}



\submitto{\NJP}

\section{Introduction}

Entanglement is of interest not only as a basic feature of quantum theory but also for applications in quantum optics including quantum cryptography \cite{cryptref}, quantum imaging \cite{qimref}, quantum spectroscopy \cite{qspecref}, and metrology \cite{metref}, among others. The strong correlations implied by entanglement have long been of interest in connection with quantum interference effects \cite{qint} and various counterintuitive experimental results, and there have been many investigations of entanglement with suitable experiments and associated theoretical analyses.

Entanglement contains some of the surprising features of quantum theory as it demonstrates a correlation stronger than is possible in classical physics and often results in anti-intuitive effects of ``non-local interactions." Different approaches were made in the past, such as the introduction of hidden parameters, to explain these observations. But it was shown that such hidden parameters are inconsistent with quantum theory and experiment. As we show in this paper, the explicit consideration of vacuum fields allows a very intuitive picture for describing quantum-optical effects that are typically explained by entanglement. 

Spontaneous parametric down-conversion (SPDC) has been one of the most useful tools in quantum-optical investigations of entanglement. In this paper we revisit SPDC from a perspective somewhat different from  much of the previous work on the subject. For this purpose we investigated the interference of two biphotons generated in two parallel, pumped down-conversion crystals as in Reference \cite{hmm}, but with no further coupling as in References \cite{mandel89,ou}. While our experimental results are in complete agreement with the earlier work, it seems worthwhile to describe and analyze them from the perspectives taken in our recently reported work \cite{heuerprl}, and to formulate in more detail the analyses outlined in that work. 

In analyzing our earlier experiments we considered in particular the role of vacuum fields and complementarity in different SPDC setups. SPDC is especially interesting in this regard because the down-converted photons result from nonlinear mixing of the pump field with the vacuum fields at the crystals, and therefore the statistical properties of the generated signal and idler fields reflect those of the vacuum fields involved in the three-wave mixing process. It has been demonstrated that, in spite of the random character of these vacuum fields, a phase memory effect could be observed if the same vacuum mode takes part in the down-conversion in two crystals \cite{ou,hrm}. Using an extended version of this experiment with three down-conversion crystals, it was shown that the absence of correlations between the vacuum fields at different crystals results in an incoherent background photon count. However, this effect of the different vacuum field modes taking part in the down-conversion could be completely overwritten, and almost perfect coherence could be observed, in the stimulated process with coherent laser radiation in these modes \cite{hmm}.

We begin in the following section by describing the experiment with two down-conversion crystals. In this experiment there are a total of four open vacuum entrances, two at each crystal, involved in the three-wave-mixing processes generating the two down-converted biphotons. No first-order coherence is observed between single-photon signal and idler fields, but high-visibility interference is observed when photons are counted in coincidence, consistent with complementarity and the absence of which-path information (Section \ref{sec:results}). In Section \ref{sec:theory} a Heisenberg-picture analysis of these results is given. In this description there is no first-order coherence observed between single-photon signal and idler fields because the vacuum fields at the four open entrances are uncorrelated, whereas second-order, ``intensity" correlations of these vacuum fields result in the observed interference effect when photons are counted in coincidence. The same formalism is then used in Section \ref{sec:hom}  to describe the Hong-Ou-Mandel (HOM) dip \cite{hom}, where we show that this consequence of entanglement and quantum interference  can again be explained equivalently in terms of non-vanishing intensity correlations of the vacuum fields. The vacuum signal and idler fields in both the two-crystal experiment and the HOM effect make explicit contributions not only to the generation of the signal and idler photons, but also to the photon coincidence counting rate. Our results are summarized and discussed further in Section \ref{sec:conc}.

\section{Experiment}\label{sec:expt}
The experimental setup consists of two crystals (BBO1 and BBO2) for the generation of down-converted signal and idler photon pairs. The crystals had a length of 4 mm and were arranged in parallel. As shown in Fig. 1, the signal channel s1 of BBO1 is overlaid with the signal channel s2 of BBO2 at the beam splitter BS1, and the idler channel i1 is overlaid with the idler channel i2 at the beam splitter BS2.
 
\begin{figure}
\centering
\includegraphics[width=8 cm]{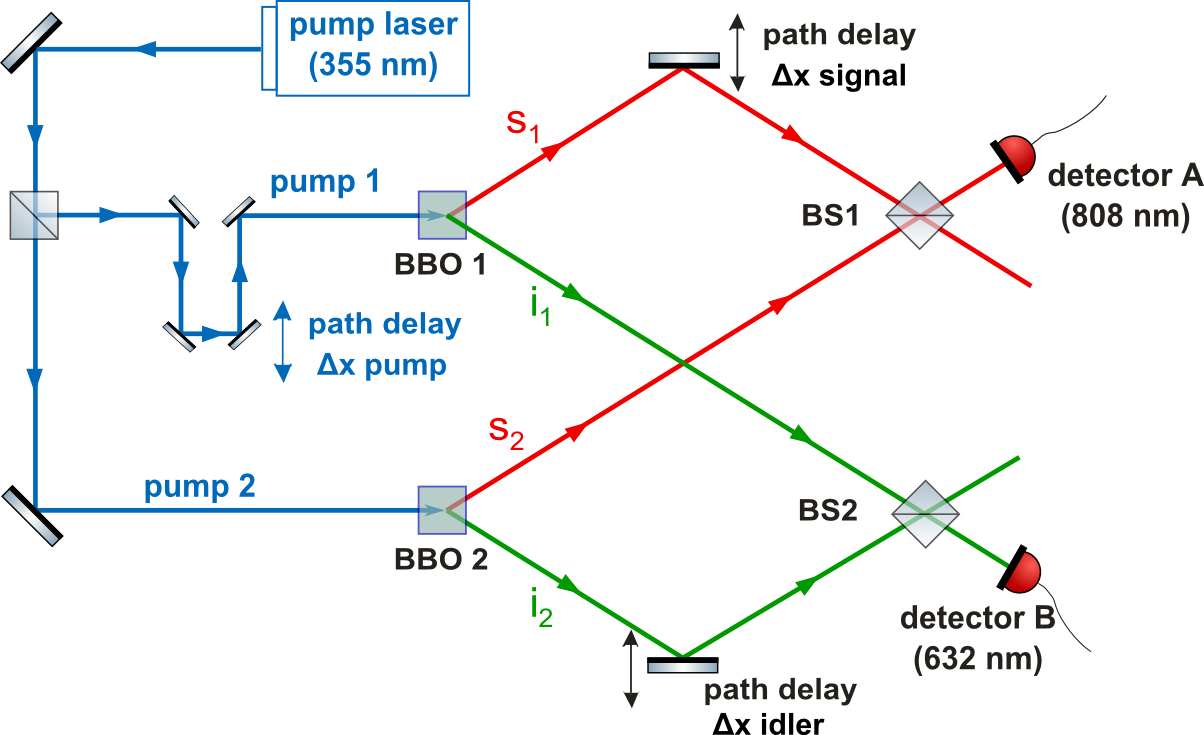}
\caption{Two-photon interference setup with two synchronously pumped SPDC crystals BBO1 and BBO2. The rate of counting signal and idler photons in coincidence at detectors A and B is measured as a function of the path delay of the pump field. The pump intensity is sufficiently weak that with high probability only one signal-idler photon pair is present during a measuring period.}
\label{fig:fig1}
\end{figure} 

The single signal photons were counted at the peak wavelength of 808 nm using a filter with a bandwidth (FWHM) of 2 nm, and the idler photons were counted at the peak wavelength of 632 nm with a bandwidth of 3 nm. The two cw, 355 nm nearly diffraction-limited pump fields from the laser (Genesis, Coherent) had a bandwidth of about 45 GHz, corresponding to a coherence length of about 1.4 mm, and were synchronized and possibly delayed with a delay line in front of BBO1. For varying the length of the signal and the idler single-photon interferometers the 100\% reflecting mirrors could be moved, resulting in phase shifts for the signal fields (upper mirror in Fig.1) and for the idler fields (lower mirror in Fig.1). Photons were counted with single-mode, fiber-coupled photodiodes (SPCM-AQRH-13, Perkin-Elmer) at the positions A and B. The whole setup was aligned with an EMCCD camera (Andor IXUS) at positions A and B. The pump power at each crystal was about 30 mW. Photon count rates at the detectors A and B were 25,000 photons/sec from crystal BBO1 and 26,000 photons/sec from BBO2. With high probability just one photon pair was in the apparatus during the measurement interval of 2 ns, so that stimulated down-conversion could be neglected in these experiments. All the observed interference effects were based on the coincidence counting of a single photon pair, either from BBO1 or from BBO2.

\section{Results}\label{sec:results}
In the first experiment the coincidence count rate was measured as a function of the signal path delay. The result is shown in Fig. 2.

\begin{figure}
\centering
\includegraphics[width=8 cm]{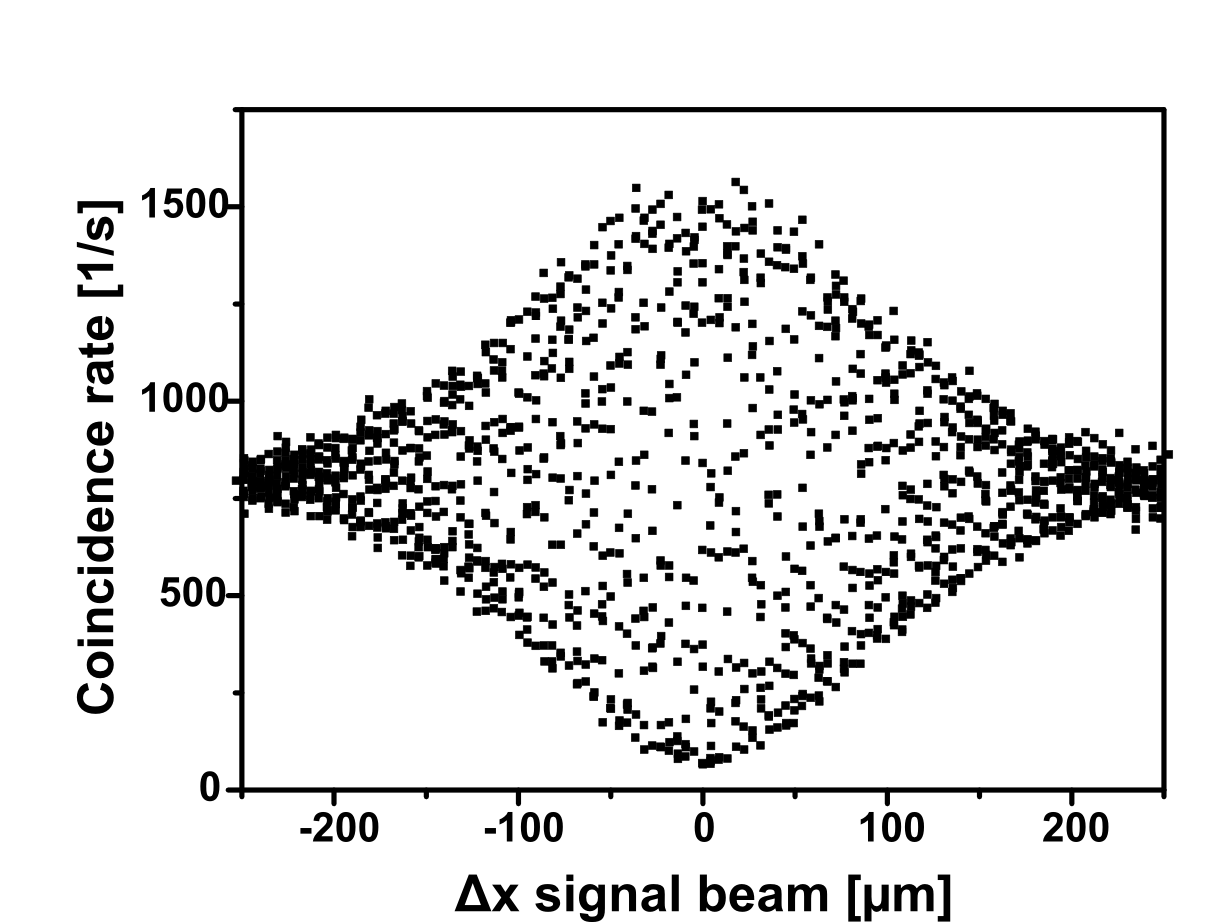}
\caption{Counting rate for single signal photons at detector A in coincidence with single idler photons at detector B as a function of the signal path delay (see Fig. 1)}.
\label{fig:fig2}
\end{figure}
 
\begin{figure}
\centering
\includegraphics[width=8 cm]{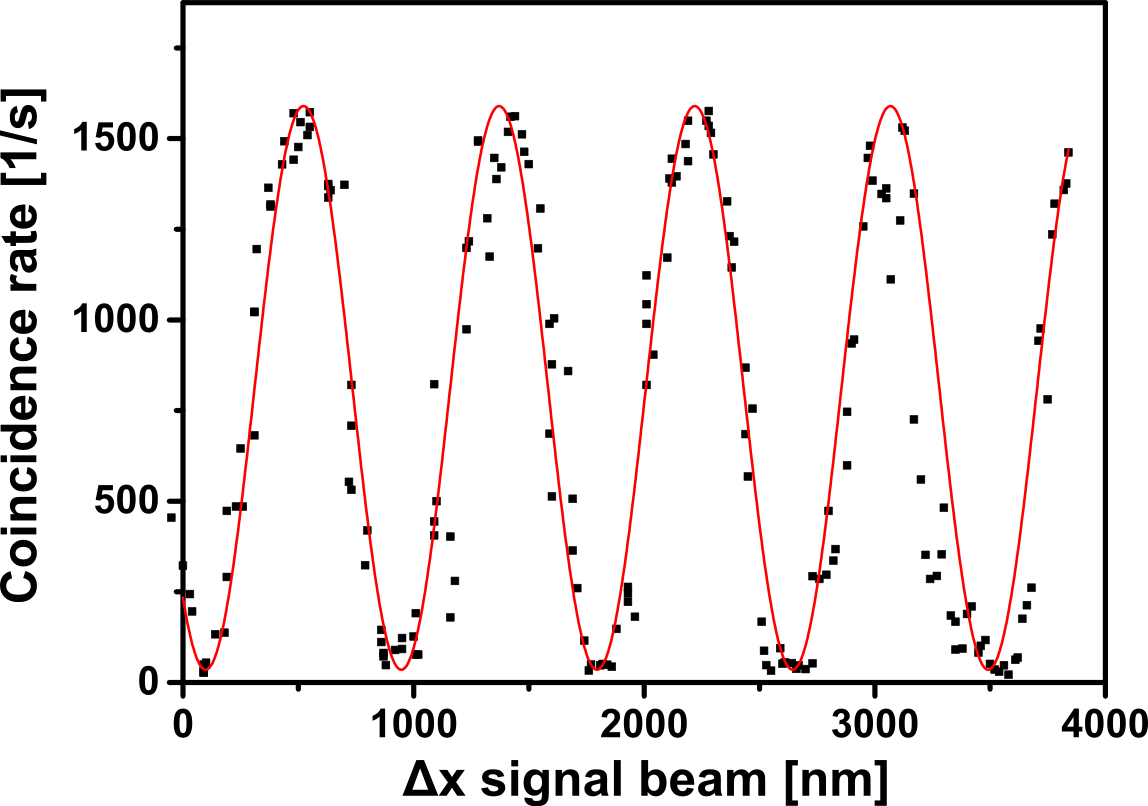}
\caption{Counting rate for single signal photons at detector A in coincidence with single idler photons at detector B as a function of the signal path delay as in Fig. 2, but with higher resolution.}
\label{fig:fig3}
\end{figure}

This measurement was done over a wide range of delays in order to determine the coherence length of the fields. From Fig. \ref{fig:fig2} the coherence length of the signal field can be estimated to be 
80 $\mu$m, in good agreement with the bandwidth of the spectral filter; the coherence length of one of the biphoton fields determines the coherence length in the second-order interference measurement. 
In a second measurement the spatial resolution of the delay line was increased in order to exhibit the biphoton interference in more detail, as shown in Fig. \ref{fig:fig3}.

The ``fringe" spacing in this biphoton interference measurement is in very good agreement with the 808-nm wavelength of the signal field. From the data shown in Fig. \ref{fig:fig3} the maximum visibility at zero delay is 94\%. Similar results are obtained when the path delay for the idler photons is varied. In that case the fringe spacing in the biphoton interference is in agreement with the idler photon wavelength of 632 nm. Since both crystals are ``coupled" only by the pump laser radiation, the observed biphoton interference with visibility $\cong 1$  may seem surprising. 

\begin{figure}
\centering
\includegraphics[width=8 cm]{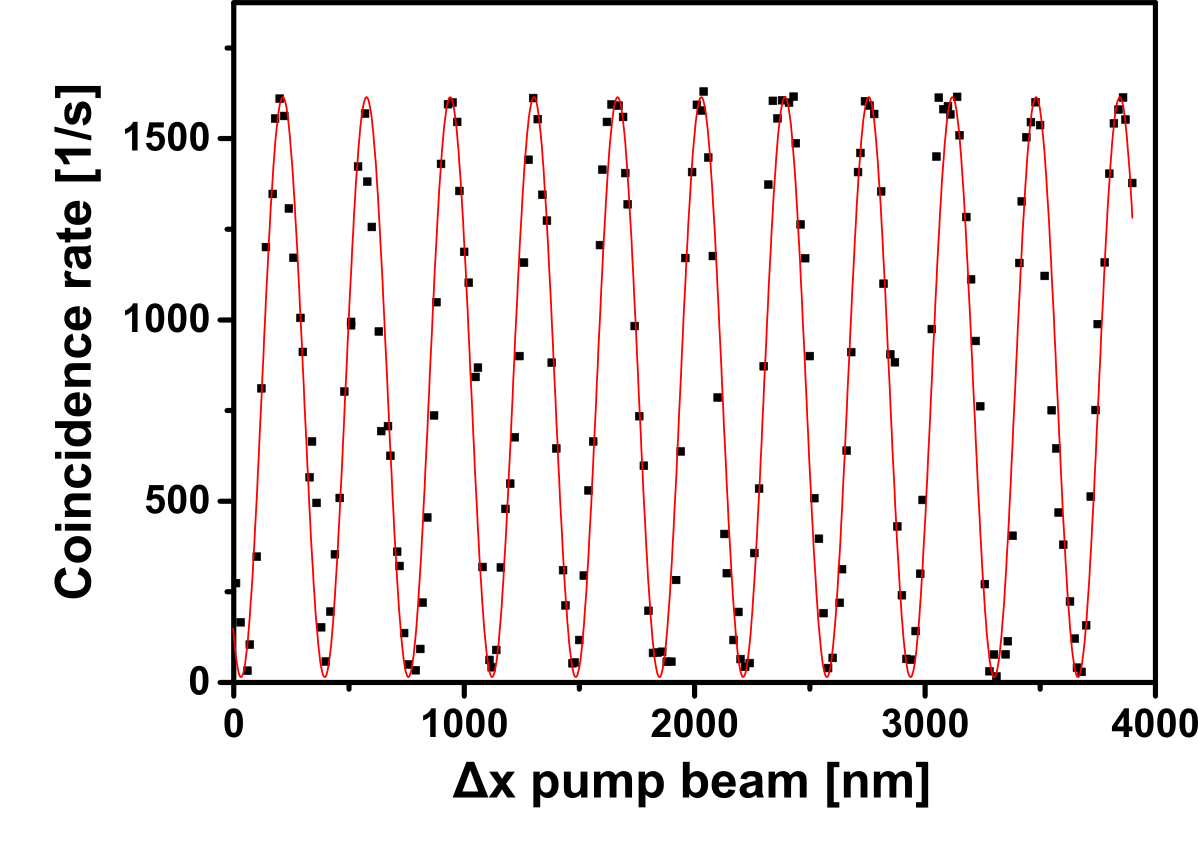}	
\caption{Second-order interference of the single signal photons from crystals BBO1 or BBO2 at detector A in coincidence with the single idler photons from crystals BBO1 or BBO2 at detector B, as a function of the pump 1 path delay for BBO1 (see Fig. \ref{fig:fig1}).}
\label{fig:fig4}
\end{figure}

The influence of the pump is shown by also varying the pump 1 path delay (Fig. \ref{fig:fig4}). It is seen that the visibility of the biphoton interference is now even higher, with a value of V=98\%. The fringe spacing of 355 nm is the wavelength of the pump laser, confirming the phase memory effect \cite{hrm} in this measurement. The coherence length evaluated from this type of measurement, over a wide range of path delays, is the coherence length of the pump laser, 1.5 mm. 

In summary, the second-order interference measurement of the biphotons emitted from two parallel, pumped and not otherwise coupled down-conversion crystals BBO1 and BBO2 shows a fringe visibility $\cong 1$. The fringe spacing observed in these measurements always corresponds to the wavelength of the delayed field.
 
\section{Theoretical Description}\label{sec:theory}
The phenomenological interaction Hamiltonian describing the downconversion process in a single crystal is 
\be
H_I=Ca_s^{\dag}a_i^{\dag}e^{-i\om_pt}+C^*a_sa_ie^{i\om_pt}
\label{ham1}
\ee
when the pump is treated as a classically prescribed, undepleted and spatially uniform field of frequency $\om_p$. $a_s$ and $a_i$ are the photon annihilation operators for the signal and idler fields satisfying the energy conservation ($\om_p=\om_s+\om_i$) and phase-matching (${\bf k}_p={\bf k}_s+{\bf k}_i$) conditions, and $C$ is proportional to the pump amplitude and the nonlinear susceptibility. To lowest order in $C$, $H_I$ in spontaneous parametric down-conversion generates an entangled field state \cite{horne,ou} that is a linear superposition of the vacuum state $|0_s0_i\ra$ and the two-photon state $|1_s1_i\ra$ consisting of a single signal photon and a single idler photon. In the Heisenberg picture the positive-frequency (photon annihilation) parts of the signal and idler electric field operators may be expressed, to lowest order in $C$, as
\be
E_s^{(+)}(\bfr,t)=a_s(t)U_s(\bfr)\cong[a_{s0}+D\ad_{i0}]U_s(\bfr)e^{-i\om_st},
\label{eqs}
\ee
and
\be
E_i^{(+)}(\bfr,t)=a_i(t)U_i(\bfr)\cong[a_{i0}+D\ad_{s0}]U_i(\bfr)e^{-i\om_it}.
\label{eqi}
\ee
$a_{s0}$ and $a_{i0}$ are the photon annihilation operators for the signal and idler fields, respectively, in the absence of the pump field. The functions $U_s(\bfr)$ and $U_i(\bfr)$ include the spatial dependence of the corresponding electric fields, which are polarized according to type I phase matching.
$D$, which is proportional to $C$, is determined by the solution of the Heisenberg equations of motion that follow from the Hamiltonian with interaction (\ref{ham1}). The specific forms of $U_s(\bfr)$, $U_i(\bfr)$, and $D$ will not be needed for our purposes.

For the two-crystal configuration of interest here,
\be
H_I=[C_1a_{s1}^{\dag}a_{i1}^{\dag}+C_2a_{s2}^{\dag}a_{i2}^{\dag}]e^{-i\om_pt} + {\rm hermitian \ conjugate},
\label{ham}
\ee
where the subscripts s1, s2, i1, and i2 refer to the modes indicated in Fig. 1. The two crystals are assumed to be identical, in which case $C_1$ and $C_2$ are proportional to the complex amplitudes of the pump fields at BBO1 and BBO2, respectively. 

The coincidence rate may be obtained simply from the superposition of probability amplitudes. Suppose first that $C_1=C_2\equiv C$. There are two (indistinguishable) ways by which a signal photon can be counted at A in coincidence with an idler photon at B. Referring to Fig. 1, these are: (i) a signal photon s1 is counted at A and an idler photon i1 at B, and (ii) a signal photon s2 is counted at A and an idler photon i2 is counted at B. In process (i), the signal field undergoes a reflection and a phase delay $\phi_1$, and the idler field is transmitted through the lower beam splitter in Fig. 1. The amplitude for the path (i) is therefore proportional to $re^{i\phi_1}t$, where $r$ and $t$ are respectively the amplitude reflection and transmission coefficients of the beam splitters, which are assumed to be identical. In process (ii), the signal field is transmitted through the upper beam splitter in Fig. 1, the idler field undergoes a reflection and a phase delay $\phi_2$, and the probability for this path is therefore proportional to $tre^{i\phi_2}$. The total probability amplitude for a signal photon to be counted at A in coincidence with an idler photon at B is therefore proportional to $rt[Ce^{i\phi_1}+Ce^{i\phi_2}]$, and the coincidence counting rate is proportional to $|rtC|^2|e^{i\phi_1}+e^{i\phi_2}|^2=2|rtC|^2(1+\cos\theta)$, $\theta=\phi_1-\phi_2$. This implies the variation of the measured coincidence rate with signal path delay shown in Fig. 3, with the oscillation period determined by the signal path delay 
($\phi_1$) and therefore the 808-nm signal wavelength.

If $C_1\neq C_2$, which would be the case if, for example, the pump amplitudes at the two crystals were different, the
coincidence rate $R_{AB}$ as discussed above will be proportional to $|rt|^2|C_1e^{i\phi_1}+C_2e^{i\phi_2}|^2$. Writing $C_2=\alpha C_1=|\alpha|e^{i\beta}$, we have
\be
R_{AB}\propto |rtC_1|^2[1+|\alpha|^2+2|\alpha|\cos\beta],
\ee
and a coincidence-rate ``visibility"
\be
V=\frac{R_{AB}^{\rm max}-R_{AB}^{\rm min}}{R_{AB}^{\rm max}+R_{AB}^{\rm min}}=\frac{2|\alpha|}{1+|\alpha|^2}.
\label{veq}
\ee
We similarly define a contrast, or ``distinguishability,"
\be
K\equiv \frac{|C_1|^2-|C_2|^2}{|C_1|^2+|C_2|^2}=\frac{1-|\alpha|^2}{1+|\alpha|^2}.
\label{keq}
\ee
It follows that 
\be
K^2+V^2=1,
\label{compeq}
\ee
which is a special case of the more general expression $K^2+V^2\le 1$ of 
complementarity \cite{remark1}.

First-order perturbation theory with the interaction Hamiltonian (\ref{ham}), assuming an initial state 
$|0_{s1}0_{i1}\ra|0_{s2}0_{i2}\ra$ with no photons in the modes $s1,i1,s2,i2$ (Fig. \ref{fig:fig1}), gives
\bea
|\psi\ra&=&|0_{s1}0{_i1}\ra|0_{s2}0_{i2}\ra+K\Big[C_1|1_{s1}1_{i1}\ra|0_{s2}0_{i2}\ra\nonumber\\
&&\mbox{}+C_2|0_{s1}0_{i1}\ra|1_{s2}1_{i2}\ra\Big].
\label{ent1}
\eea
Here $|1_{s1}1_{i1}\ra$ is the state with one photon in each of the modes $s1$ and $i1$, etc. The factor $K$ need not be specified, nor will it be necessary to normalize  $|\psi(t)\ra$ for our purposes. As discussed in
much detail in Reference \cite{mandel89}, the ``entanglement with the vacuum" described by the state 
(\ref{ent1}) implies the observed two-photon coincidence rate. In terms of interfering probability amplitudes as discussed above, the amplitude to count (annihilate) a signal photon $s1$ at detector A in coincidence with an idler photon $i1$ at detector B, including the phase delay $\phi_1$, is proportional to $\la{\rm vac}|a_{s1}e^{i\phi_1}a_{i1}|\psi\ra=C_1e^{i\phi_1}$, where the vacuum state of the field is denoted by
\be
|{\rm vac}\ra=|0_{s1}0_{i1}\ra|0_{s2}0_{i2}\ra.
\label{vacstate}
\ee
Similarly the amplitude to count a signal photon $s2$ at detector A in coincidence with an idler photon $i2$ at detector B, is proportional to $\la{\rm vac}|a_{s2}a_{i2}e^{i\phi_2}|\psi\ra=C_2e^{i\phi_2}$, and the
rate for counting a signal photon at A in coincidence with an idler photon at B is therefore proportional
to $|C_1e^{i\phi_1}+C_2e^{i\phi_2}|^2$. This simple argument emphasizes the role played by entanglement with the vacuum field: to explain the observed coincidence rate we must include the vacuum states $|0_{s1}0_{i1}\ra$ and $|0_{s2}0_{i2}\ra$ as well as the two-photon states $|1_{s1}1_{i1}\ra$ and
$|1_{s2}1_{i2}\ra$ in describing the state of the field.

We now turn to a simplified Heisenberg-picture analysis that exhibits explicitly the role of the vacuum {\sl fields}, as opposed to the vacuum {\sl states}, participating in the downconversion at the two crystals. The rate for counting a signal photon at detector A in coincidence with an idler photon at detector B may be taken to be proportional to \cite{glauber} 
\be
R_{AB}=\la{\rm vac}| E_A^{(-)}E_B^{(-)}E_B^{(+)}E_A^{(+)}|{\rm vac}\ra,
\label{eq99}
\ee
where $E_A^{(+)}$ and $E_B^{(+)}$ are the positive-frequency parts of the electric field operators at detectors A and B, $E_A^{(-)}$, $E_B^{(-)}$ are their hermitian conjugates, and $|{\rm vac}\ra$ is again the vacuum field state. As in the preceding discussion we treat A and B as effectively point detectors, since as a practical matter the spatial integrations over detector volumes only introduces a multiplicative factor of no interest for our purposes. We therefore ignore entirely the spatial variations of the electric fields, except of course for the path delays described by $\phi_1$ and $\phi_2$. Then to calculate the dependence of $R_{AB}$ on these path delays we can effectively take, to first order in $D$,
\be
E_A^{(+)}=\Big([a_{s10}+D\ad_{i10}]re^{i\phi_1}+[a_{s20}+D\ad_{i20}]t\Big)e^{-i\om_st}
\label{eq100}
\ee
and
\be
E_B^{(+)}=\Big([a_{i10}+D\ad_{s10}]t+[a_{i20}+D\ad_{s20}]re^{i\phi_2}\Big)e^{-i\om_it},
\label{eq101}
\ee
which are obvious extensions of (\ref{eqs}) and (\ref{eqi}). 

Suppose first that only the crystal BBO1 is pumped. Since $a_{s10}|{\rm vac}\ra=a_{s20}|{\rm vac}\ra=a_{i10}|{\rm vac}\ra=a_{i20}|{\rm vac}\ra=0$, and the photon operators for different modes commute, it follows that
\be
\la E_B^{(-)}E_A^{(+)}\ra=0,
\ee
i.e., the signal and idler fields in SPDC are uncorrelated \cite{hmm}. The signal-idler photon coincidence rate is found similarly; to second order in D,
\be
R_{AB}\propto |D|^2|rt|^2\la{\rm vac}|a_{i10}\ad_{i10}a_{i10}\ad_{i10}|{\rm vac}\ra=|D|^2|rt|^2.
\label{eq104}
\ee
The dependence of this signal-idler coincidence rate on only the vacuum idler field reflects the fact that a signal photon at detector A, produced by the mixing of the vacuum idler field with the pump field at BBO1, is accompanied by its partner idler photon at detector B. Since $[E_A^{(+)},E_B^{(+)}]=0$, we can also write $R_{AB}$ in the form
\bea
R_{AB}&\propto&\la{\rm vac}| E_B^{(-)}E_A^{(-)}E_A^{(+)}E_B^{(+)}|{\rm vac}\ra\nonumber\\
&=&|D|^2|rt|^2\la{\rm vac}|a_{s10}\ad_{s10}a_{s10}\ad_{s10}|{\rm vac}\ra\nonumber\\
&=&|D|^2|rt|^2,
\label{eq1004}
\eea
which can be interpreted in the same way as (\ref{eq104}) except that now the roles of the signal and idler fields are switched. The fact that we can write $R_{AB}$ in terms of either signal or idler vacuum expectation values simply reflects the fact that signal and idler photons are generated in pairs.

For the experiment of Fig. 1 in which both crystals are pumped we obtain, from (\ref{eq99})-(\ref{eq101}),
\bea
R_{AB}&\propto& |D|^2|rt|^2\Big[\la{\rm vac}|a_{i10}\ad_{i10}a_{i10}\ad_{i10}|{\rm vac}\ra \nonumber\\
&&\mbox{}+\la{\rm vac}|a_{i20}\ad_{i20}a_{i20}\ad_{i20}|{\rm vac}\ra \nonumber\\
&&\mbox{}+2\la{\rm vac}|a_{i10}\ad_{i10}a_{i20}\ad_{i20}|{\rm vac}\ra\cos\theta\Big]
\label{eq102}
\eea
to lowest order in $D$, where again $\theta=\phi_1-\phi_2$. Equivalently, since $[E_A^{(+)},E_B^{(+)}]=0$,
\bea
R_{AB}&\propto& |D|^2|rt|^2\Big[\la{\rm vac}|a_{s10}\ad_{s10}a_{s10}\ad_{s10}|{\rm vac}\ra \nonumber\\
&&\mbox{}+\la{\rm vac}|a_{s20}\ad_{s20}a_{s20}\ad_{s20}|{\rm vac}\ra \nonumber\\
&&\mbox{}+2\la{\rm vac}|a_{s10}\ad_{s10}a_{s20}\ad_{s20}|{\rm vac}\ra\cos\theta\Big].
\label{eq102a}
\eea
Since all the vacuum expectation values in (\ref{eq102}) and (\ref{eq102a}) $=1$, we obtain again 
\be
R_{AB}\propto 1+\cos\theta.
\ee
The generalization to non-identical crystals or pump fields is straightforward and of course reproduces the complementarity relation (\ref{compeq}). 

In the derivation of $R_{AB}$ based on probability amplitudes the $\cos\theta$ term resulted from the interference of the amplitudes for the two indistinguishable processes (i) and (ii). Obviously the interference here is similar to the interference of two probability amplitudes in the two-slit experiment with single photons, with the difference that $R_{AB}$ describes the counting of {\sl two} photons (signal and idler) in coincidence. The complementarity relation (\ref{compeq}) applies to both examples. In the classical description of the two-slit experiment, for example, the visibility of interference fringes is reduced when the difference in the field intensities incident on the two slits is increased. In the experiment of Fig. 1 the 
visibility (\ref{veq}) is reduced when the distinguishability of the processes (i) and (ii) is increased.  

In the Heisenberg-picture calculation the role of the vacuum fields taking part in SPDC is explicit in the expression (\ref{eq102}). The vacuum fields incident on BBO1 and BBO2 are associated with $a_{i10}$ and $a_{i20}$, respectively, and the last term on the right-hand side of (\ref{eq102}) is nonvanishing because these fields have quantum fluctuations and zero-point energy, e.g., $\la{\rm vac}|\frac{1}{2}\hbar\om_ia_{i10}\ad_{i10}|{\rm vac}\ra=\frac{1}{2}\hbar\om_i$. 

Given that the coincidence rate can be understood in terms of a state describing entanglement with the vacuum, it is not surprising that in the Heisenberg picture the role of the vacuum field is exhibited explicitly in the expression for $R_{AB}$. Perhaps more interesting, however, is the fact that the contribution of the vacuum field to $R_{AB}$ involves more than the fact that it mixes with the pump field to generate biphotons at the crystal. It also involves propagation to the detector of the vacuum fields in the biphoton modes, independent of their mixing with the pump. We can express (\ref{eq1004}), for example, as
\be
R_{AB}\propto\la{\rm vac}| E_{Ds1}^{(-)}E_{Vi1}^{(-)}E_{Vi1}^{(+)}E_{Ds1}^{(+)}|{\rm vac}\ra,
\label{eq1005}
\ee
where 
\be
E_{Ds1}^{(+)}\propto Da_{i10}^{\dag}re^{i\phi_1}
\ee
is the part of the positive-frequency signal field in mode $s1$ that is {\sl generated} in the down-conversion process and propagated to detector A, and 
\be
E_{Vi1}^{(+)}\propto ta_{i10}
\ee
is the positive-frequency part of the {\sl vacuum} idler field in mode $i1$ and propagated to detector B. Thus
the coincidence rate (\ref{eq1005}) can be interpreted as a coincidence between a signal photon at A
and an idler vacuum field fluctuation at B.

\section{Vacuum Fields and the Hong-Ou-Mandel Dip}\label{sec:hom}
The same simplified formalism in which the role of the vacuum fields is explicit can be used to describe the 
Hong-Ou-Mandel-dip \cite{hom} (Fig. 5). In this case photons can be counted in coincidence at detectors A and B if (i) both signal and idler fields are reflected by the beam splitter BS, or (ii) both signal and idler fields are transmitted by BS. The probability amplitudes for these indistinguishable processes are proportional to $r^2$ and $t^2$, respectively. The coincidence rate is therefore proportional to $|r^2+t^2|^2$, and a dip in the coincidence rate ($R_{AB}=0$) is then obtained when $|r|=|t|=1/\sqrt{2}$, since $r=te^{i\pi/2}$.

\begin{figure}
\centering
\includegraphics[width=8 cm]{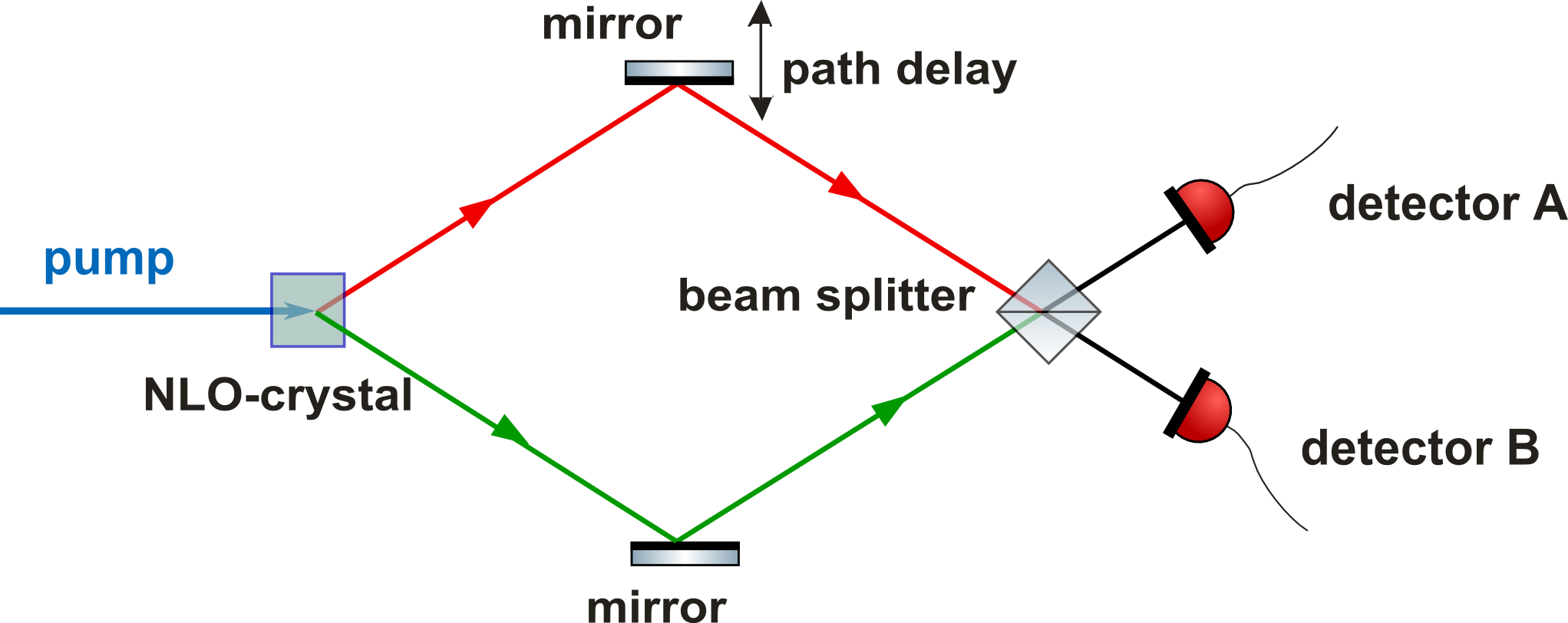}	
\caption{Hong-Ou-Mandel setup for observation of a dip in the signal-idler photon coincidence rate at detectors A and B. The signal and idler fields, which are assumed to have the same frequency and polarization, are reflected by the upper and lower mirrors, respectively, and can either be reflected or transmitted by the beam splitter.}
\label{fig:fig5}
\end{figure}

We can also relate the HOM interference effect to ``entanglement with the vacuum" by writing the field state
given by perturbation theory as 
\bea 
|\psi\ra&=&|0_s0_i\ra_A|0_s0_i\ra_B+Dt^2|1_s1_i\ra_A|0_s0_i\ra_B\nonumber\\
&&\mbox{}+Dr^2|0_s0_i\ra_A|1_s1_i\ra_B,
\eea
similarly to (\ref{ent1}), where now A and B refer to field modes incident at A and B, and of course the transition amplitudes implied by this state give $R_{AB}\propto |r^2+t^2|^2$.

In our simplified Heisenberg-picture approach the positive-frequency parts of the electric field operators at the detectors A and B are
\be
E_A^{(+)}=\Big[r(a_{s0}+D\ad_{i0})e^{-i\om_st}+t(a_{i0}+D\ad_{s0})e^{-i\om_it}\Big]
\label{hom100}
\ee
and
\be
E_B^{(+)}=\Big[t(a_{s0}+D\ad_{i0})e^{-i\om_st}+r(a_{i0}+D\ad_{s0})e^{-i\om_it}\Big],
\label{hom101}
\ee
where the signal and idler fields are assumed to traverse the same distance. Evaluating $R_{AB}=\la{\rm vac}| E_A^{(-)}E_B^{(-)}E_B^{(+)}E_A^{(+)}|{\rm vac}\ra$ in the same way as for the two-crystal experiment, we obtain, to lowest order in $D$,
\bea
R_{AB}&=&\Big[|t|^4\la{\rm vac}|a_{s0}\ad_{s0}a_{s0}\ad_{s0}|{\rm vac}\ra\nonumber\\
&&\mbox{}+|r|^4\la{\rm vac}|a_{i0}\ad_{i0}a_{i0}\ad_{i0}|{\rm vac}\ra\nonumber\\
&&\mbox{}+r^2t^{*2}\la{\rm vac}|a_{s0}\ad_{s0}a_{i0}\ad_{i0}|{\rm vac}\ra\nonumber\\
&&\mbox{}+r^{*2}t^2\la{\rm vac}|a_{i0}\ad_{i0}a_{s0}\ad_{s0}|{\rm vac}\ra\Big]|D|^2\nonumber\\
&=&|r^2+t^2|^2|D|^2,
\label{hom301}
\eea
since the vacuum expectation values are all 1. Since $[E_A^{(+)},E_B^{(+)}]=0$, we can also write
\bea
R_{AB}&=&\Big[|t|^4\la{\rm vac}|a_{i0}\ad_{i0}a_{i0}\ad_{i0}|{\rm vac}\ra\nonumber\\
&&\mbox{}+|r|^4\la{\rm vac}|a_{s0}\ad_{s0}a_{s0}\ad_{s0}|{\rm vac}\ra\nonumber\\
&&\mbox{}+r^2t^{*2}\la{\rm vac}|a_{s0}\ad_{s0}a_{i0}\ad_{i0}|{\rm vac}\ra\nonumber\\
&&\mbox{}+r^{*2}t^2\la{\rm vac}|a_{i0}\ad_{i0}a_{s0}\ad_{s0}|{\rm vac}\ra\Big]|D|^2.\nonumber\\
\label{hom300}
\eea
In either form the HOM dip ($R_{AB}=0$) then follows for $|r|=|t|=1/\sqrt{2}$ and the fact that $r=te^{i\pi/2}$.  

These expressions have some similarity to those encountered in the theory of the Hanbury Brown-Twiss intensity interference of chaotic classical fields. Thus the first and second terms in (\ref{hom300}), for instance, correspond in effect to mean-square intensities, while the third and fourth terms effectively correspond to a (second-order) signal-idler ``intensity" interference and are nonvanishing even though there is no (first-order) interference of the signal and idler fields. The difference in the HOM case is that the interference is between the signal and idler {\sl vacuum} field ``intensities" proportional to 
$\la a_{s0}\ad_{s0}\ra$ and $\la a_{i0}\ad_{i0}\ra$, respectively.

\section{Summary and Concluding Remarks}\label{sec:conc}
The high-visibility interference of the biphotons from two parallel pumped but otherwise completely decoupled crystals shown in Fig. \ref{fig:fig4} can be intuitively understood with a model based on signal and idler vacuum {\sl fields}. The vacuum signal and idler fields not only mix with the pump field to generate the biphotons; the ``un-mixed" parts of these fields in the same biphoton modes also propagate and collect delays to the detectors and act there with the generated fields to give the measured coincidence counting rate [cf. Eq. (\ref{eq1005})].

This model leads to the same predictions as the so far used conventional picture emphasizing entanglement and complementarity, based on interfering probability amplitudes, consistent with complementarity and the absence of which-path information in the coincidence counting of signal photons at detector A and idler photons at detector B. This explanation emphasizes the (entangled) nature of the field state generated in the experiment of Figure \ref{fig:fig1}.

We have also shown with our simplified model that the Hong-Ou-Mandel dip can be described not only as an interference effect resulting from entanglement with the vacuum state, but also as a consequence of vacuum
{\sl fields}. As in the simplified theory of the two-crystal experiment, the expressions we obtained for the HOM interference [Eqs. (\ref{hom301}) and (\ref{hom300})] involve only vacuum fields.

The vacuum-field interpretations discussed in this paper are obviously not the only ones possible. Our intention here has only been to show that, within the context of spontaneous parametric down-conversion, the vacuum-field picture allows for a simple, alternative, and fairly intuitive approach to the understanding of results obtained in different experimental configurations. 

\ack
We acknowledge the support of Deutsche Forschungsgemeinschaft (German Research Foundation) and Open Access Publication Fund of Potsdam University\\

\end{document}